\documentclass[apjl]{emulateapj}
\usepackage{apjfonts,graphicx,hyperref,epsf}
\usepackage{subfigure}
\begin{document}

\shortauthors{Morningstar et al.}

\title{A Seyfert-2-like Spectrum in the High-Mass X-ray Binary Microquasar V4641 Sgr}

\author{Warren~R.~Morningstar\altaffilmark{1},
        Jon~M.~Miller\altaffilmark{1},
        M.~T.~Reynolds\altaffilmark{1},
        Dipankar Maitra\altaffilmark{2}}

\altaffiltext{1}{Department of Astronomy, University of Michigan, 500
  Church Street, Ann Arbor, MI 48109-1042, wmorning@umich.edu,
  jonmm@umich.edu}

\altaffiltext{2}{Department of Physics \& Astronomy, Wheaton College,
  Norton, MA, 02766, USA}


\label{firstpage}

\begin{abstract}
We present an analysis of three archival Chandra observations of the
black hole V4641~Sgr, performed during a decline into quiescence.  The
last two observations in the sequence can be modeled with a simple
power-law.  The first spectrum, however, is remarkably similar to
spectra observed in Seyfert-2 active galactic nuclei, which arise
through a combination of obscuration and reflection from distant
material.  This spectrum of V4641~Sgr can be fit extremely well with a
model including partial-covering absorption and distant reflection.
This model recovers a $\Gamma \simeq 2.0$ power-law incident spectrum,
typical of black holes at low Eddington fractions.  The implied
geometry is plausible in a high-mass X-ray binary like V4641~Sgr, and
may be as compelling as explanations invoking Doppler-split line
pairs in a jet, and/or unusual Comptonization.  We discuss potential
implications and means of testing these models.
\end{abstract}

\section{Introduction}

The nature of X-ray emission at low Eddington fractions remains
uncertain.  It is particularly interesting to explore connections
between accretion inflows and jet outflows in X-ray binaries at
$L\simeq 10^{-4}~{\rm L}_{Edd}$, as this regime may provide insights
accross the mass-scale into the much larger class of low-luminosity 
active galactic nuclei (LLAGN) which exist at comparably low Eddington 
fractions.  Recent work on V404 Cyg, for instance, has shown 
that jet production can occur over many orders of magnitude (see e.g. 
Gallo, Fender, \& Pooley\ 2003).

One particularly well-studied jet-producing X-ray binary is SS 433.
SS 433 is unique in that its jets can be resolved in different
wavelengths, removing ambiguities in how spectral signatures map onto
different parts of the accretion flow.  In particular, strong X-ray
emission lines from the extended jets strongly imply an outflow power
that exceeds the radiative Eddington limit for reasonable compact
object masses (see, e.g., Marshall et al.\ 2013).  Of course, the
mechanical luminosity is not governed by the radiative Eddington
limit, but since the power in the jets is likely generated in part via
accretion, it is possible that the mass accretion rate in SS 433 is
super-Eddington.

It has recently been suggested that V4641 Sgr may be a precessing
micro-blazar that launches extreme jets, based on an unusual {\it
  Chandra} X-ray spectrum obtained at low luminosity (Gallo, Plotkin,
\& Jonker 2014).  The spectrum shows a strange hump or break, near to
and above 4 keV.  It may require a combination of
relativistically--shifted emission lines from jets, as in the case of SS 433, 
and/or unusual continuum emission also tied to the jets.  It is notable that 
the feature(s) detected in V4641 Sgr are far stronger than lines detected
in SS 433.

However, in any high-mass X-ray binary (HMXB), the massive companion
wind may serve to complicate inferences regarding the accretion flow.
Such winds can be clumpy, and can give rise to aperiodic obscuration
that may fully or partially cover the central engine (the well-known
dips in Cygnus X-1 are one prominent example of clumps; see e.g. Hanke et 
al.\ 2009 ).  Such obscuration has been observed on multiple occasions in 
the spectrum of V4641 Sgr in both optical and X-ray wavelengths (Maitra 
\& Bailyn\ 2006; Charles et al.\ 1999; Revnivtsev et al.\ 2002b).  
Even in the absence of clumps, shocks and ionization fronts can 
lead to an inhomogeneous medium (see, e.g., Watanabe et al.\ 2006).

When the direct continuum is wholly or partly blocked in a HMXB, the
equivalent width of emission lines (arising via irradiation of the
wind) can become very large (even 1--2 keV or more; see White, Nagase,
\& Parmar 1995).  The full complexity of HMXB winds has become clear
in the era of advanced CCDs and dispersive X-ray spectrometers (see,
e.g., Torrejon et al.\ 2010).  In Vela X-1, which harbors a B-type
star (albeit one with a stronger wind than V4641~Sgr), excellent 
X-ray spectra and complex wind models have been compared; a wind with 
multiple zones, shocks, and columns that can approach $N_{H} \geq 10^{23}~{\rm
  cm}^{-2}$ is clearly indicated (Watanabe et al. 2006).

V4641 Sgr is highly unusual even among HMXBs, and unique in other
respects.  The system harbors a black hole (${\rm M}_{BH} = 9.6^{+2.1}_{-0.9}~{\rm
  M}_{\odot}$) accreting from a high-mass B9 III-type
(${\rm M}_{C} = 6.5^{+1.6}_{-1.0}~{\rm M}_{\odot}$) companion star (Orosz et
al.\ 2001, although MacDonald et al. 2014 find slightly lower component masses:
${\rm M}_{BH}=6.4\pm0.6~{\rm M}_{\odot}$ and 
${\rm M}_{C}=2.9\pm0.4~{\rm M}_{\odot}$).  However, the inclination angle of 
the superluminal jet detected in 1999 by Hjellming et al. (2000) is likely $i_{jet} \leq 10^{\circ}$ , 
whereas the inclination of the binary system is likely much higher, $60^{\circ}
< i_{binary} < 70^{\circ}$ (Orosz et al.\ 2001).  It is not clear how
this geometry might affect observed spectra.


In this work, we examine {\it Chandra} observations of V4641~Sgr at
low Eddington fractions, originally analyzed by Gallo, Plotkin, \&
Jonker (2014).  We suggest an alternative explanation for the
SS~433-like features that may be present in the spectrum.  The
observations and data reduction are described in Secton 2.  In Section
3, we report the results of our analysis.  Finally we discuss our
results in section 4.

\section{Observations and Data Reduction}
The data we utilize are archived {\it Chandra} observations performed 
in 2004 as the source was declining into a quiescent state.
V4641 Sgr was observed at low flux in observations 4451 (start time:
MJD 53203.45, duration: 9.2~ks), 4452 (start time: MJD 53216.76,
duration 18.3~ks), and 4453 (MJD 53227.18, duration: 36.5 ks), using
the ACIS-S detector.  The CCD was operated in ``FAINT'' mode using a
sub-array to prevent photon pile-up distortions.  CIAO version 4.4 in
the most up-to-date calibration files were used to reduce the ``evt2''
files.  Using the ``specextract'' script, source and background
spectra and responses were created from each observation.  The source
spectra were extracted using a circle centered on the source, with a
radius of 10 pixels (4.92 arcseconds).  Background spectra were
extracted from an adjacent, source-free region of the same size.

The HEASARC/FTOOLS suite was used to further process the data.  The
familiar ``grppha'' was used to group all spectra to require 15 counts
per bin, in order to ensure the validity of $\chi^{2}$ statistics
(cash 1979).  The spectra were subsequently analyzed using XSPEC
version 12.8.0 (Arnaud 1996).  Models were fitted over the 0.5--8.0
keV energy range.

To facilitate a comparison with V4641~Sgr, we downloaded the 2007
{\it XMM-Newton} observation of the well-known Seyfert-2 NGC 7582
(observation 0405380701; see Bianchi et al.\ 2009).  The XMM data were
reduced using SAS version 13.0.  Potential background flares were
excised by analyzing the light curve of a background region on the
same chip as the source.  Standard filtering was performed
(e.g. PATTERN $\leq$ 4 and FLAG $=$ 0).  Source and background spectra
were extracted from circles with 30 arc-second radii, and
corresponding response files were made using ``rmfgen'' and
``arfgen''.  The net exposure time after processing was 15.1 ksec.
The spectra were grouped using the FTOOL ``grppha'' to require at
least 15 counts per bin, as was done for the {\it Chandra} spectra of 
V4641 Sgr.

\section{Analysis \& Results}

\begin{figure}[tb]
\begin{center}
\includegraphics[width=0.65\hsize, angle=270]{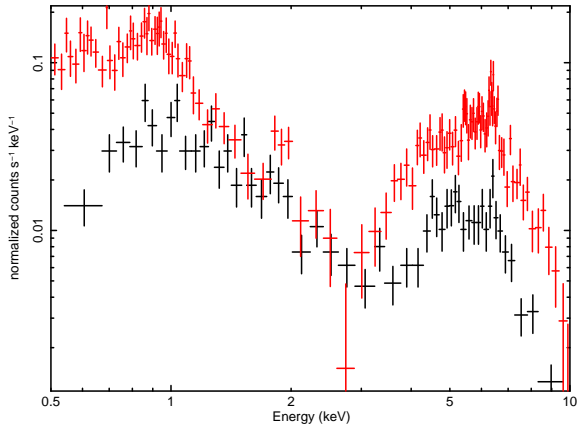}
\caption[b]{The figure above shows the {\it Chandra} 
  spectrum of V4641 Sgr (in black), and a 2007 {\it XMM-Newton} 
  observation of the well-known Seyfert-2 NGC 7582 (in red).
  Both spectra were binned to require at least 15 counts per bin; the spectrum
  of NGC 7582 was further binned for visual clarity only.  No spectral
  fitting has been performed, nor has any flux scaling.  Features
  consistent with a partical-covering absorption plus distant
  reflection geometry are evident in both spectra, and dominate
  differences between the {\it Chandra} and {\it XMM-Newton} area
  functions.}
  \vspace{0cm}
  \label{fig:combined}
\end{center}
\end{figure}

Figure~\ref{fig:combined} shows the {\it Chandra} spectrum of
observation 4451 plotted with the {\it XMM-Newton} spectrum of the
Seyfert-2 NGC 7582.  The similarities between the two spectra are
clear.  Typical of Seyfert-2 AGN, Bianchi et al.\ (2009) find that the
spectrum of NGC 7582 is fitted acceptably by a combination of strong
and complex absorption, and distant reflection.  In view of the
spectral similarities between V4641 Sgr and NGC 7582, we constructed
the following model: $tbabs\times pcfabs\times pexmon$ (Wilms, Allen
\& McCray 2000, Nandra et al. 2007).  The use of a single ``pcfabs''
component represents a simplified version of the multiple-component
absorption and/or partial-covering absorption typically required to
fit Seyfert-2 X-ray spectra (e.g. Turner et al.\ 1997; Matt et
al.\ 2000; Risaliti, Elvis, \& Nicastro 2002).  ``Pexmon''
self-consistently generates Fe K emission lines and reflection
features.

Within ``pexmon'', an inclination $i=0^{\circ}$ was fixed (however 
we find that the quality of the fit is insensitive to the inclination 
and similar values of the other parameters are returned when the binary 
inclination of $70^{\circ}$ is used), and the abundance of elements 
heavier than Helium equivalent to the solar abundance (apart from iron, 
where an abundance $A_{\rm Fe}=0.3A_{\rm  Fe, solar}$ was fixed; fits 
are only slightly worse with a solar iron abundance: 
$\chi^{2}_{\nu}=0.82$ as opposed to $\chi^{2}_{\nu}=0.77$).  The 
cutoff energy ($E_{c}$) was fixed at $10^{6}~{\rm keV}$.  The other 
five parameters; the photon index ($\Gamma$), the scaling factor for 
reflection ($f_{\rm refl}$), the normalization (K), the internal 
Hydrogen column density $N_{H}$, and its covering fraction 
$f_{\rm cov}$ were allowed to vary.  The Galactic equivalent neutral 
hydrogen column density was fixed to 
$N_{H} = 0.18\times10^{22}~{\rm atoms}~{\rm cm}^{-2}$ (Kalberla et
al.\ 2005).

\begin{figure*}[htb]
\subfigure{\includegraphics[width=0.32\hsize, angle=270]{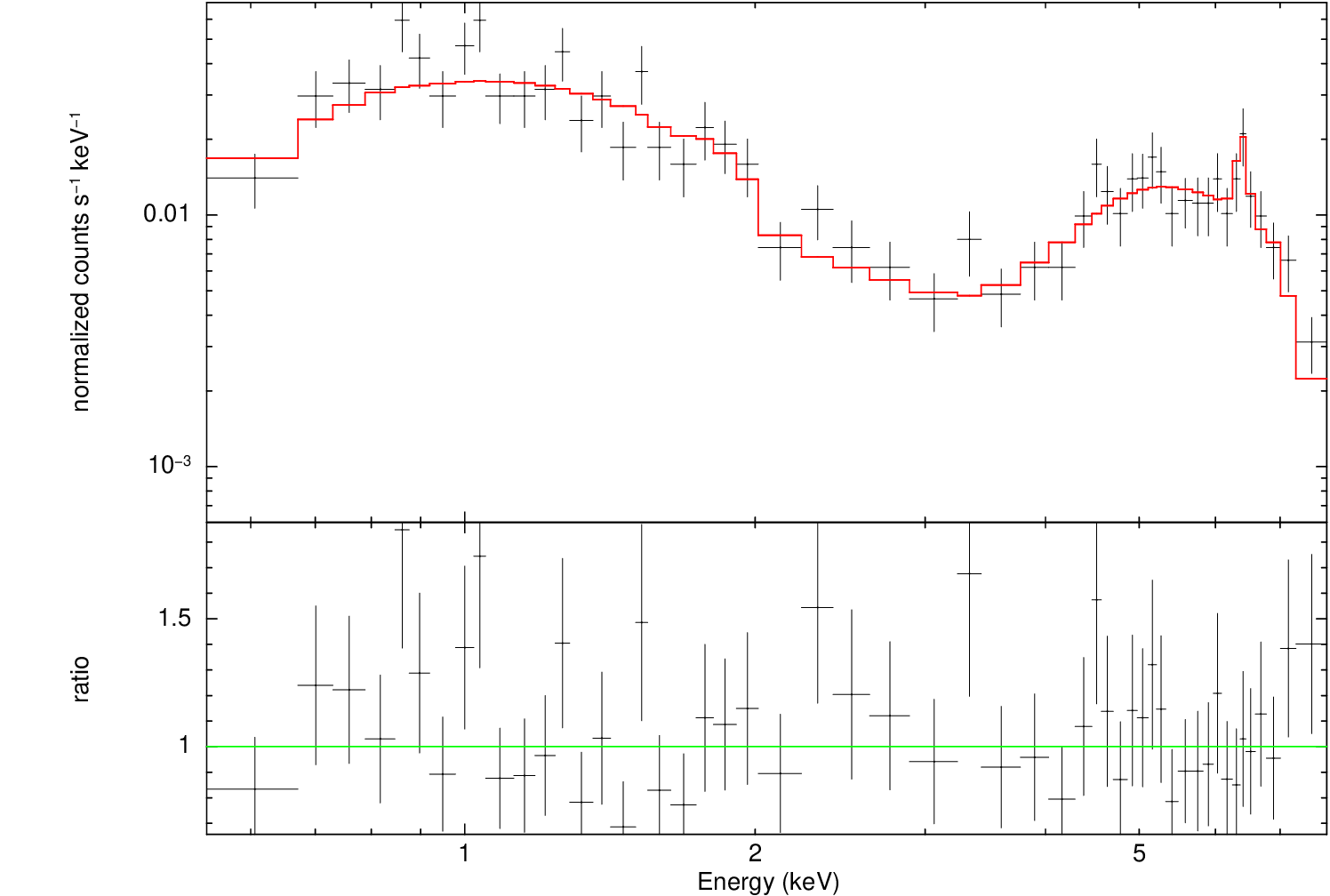}}\quad
\subfigure{\includegraphics[width=0.32\hsize, angle=270]{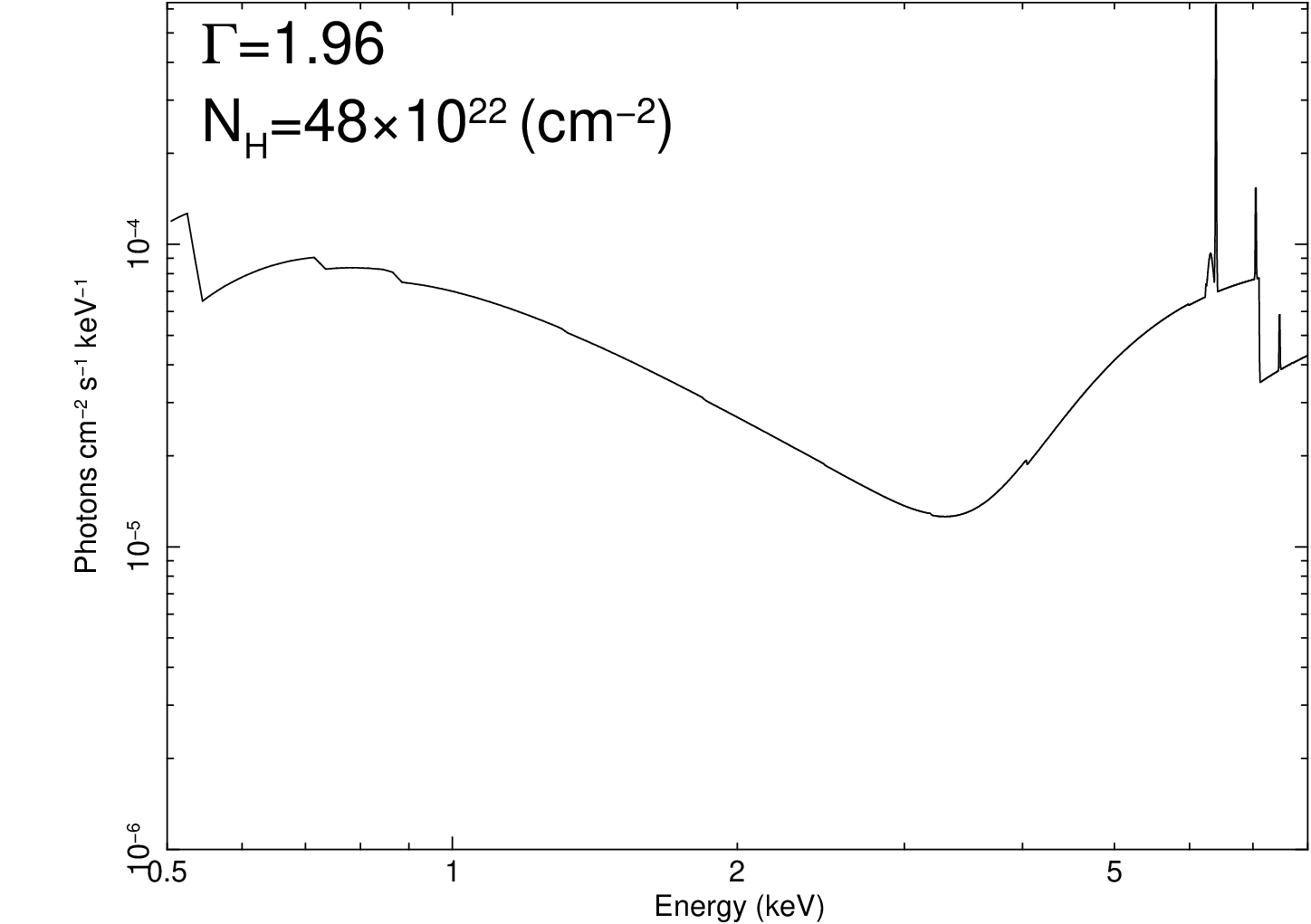}}\quad
\subfigure{\includegraphics[width=0.32\hsize, angle=270]{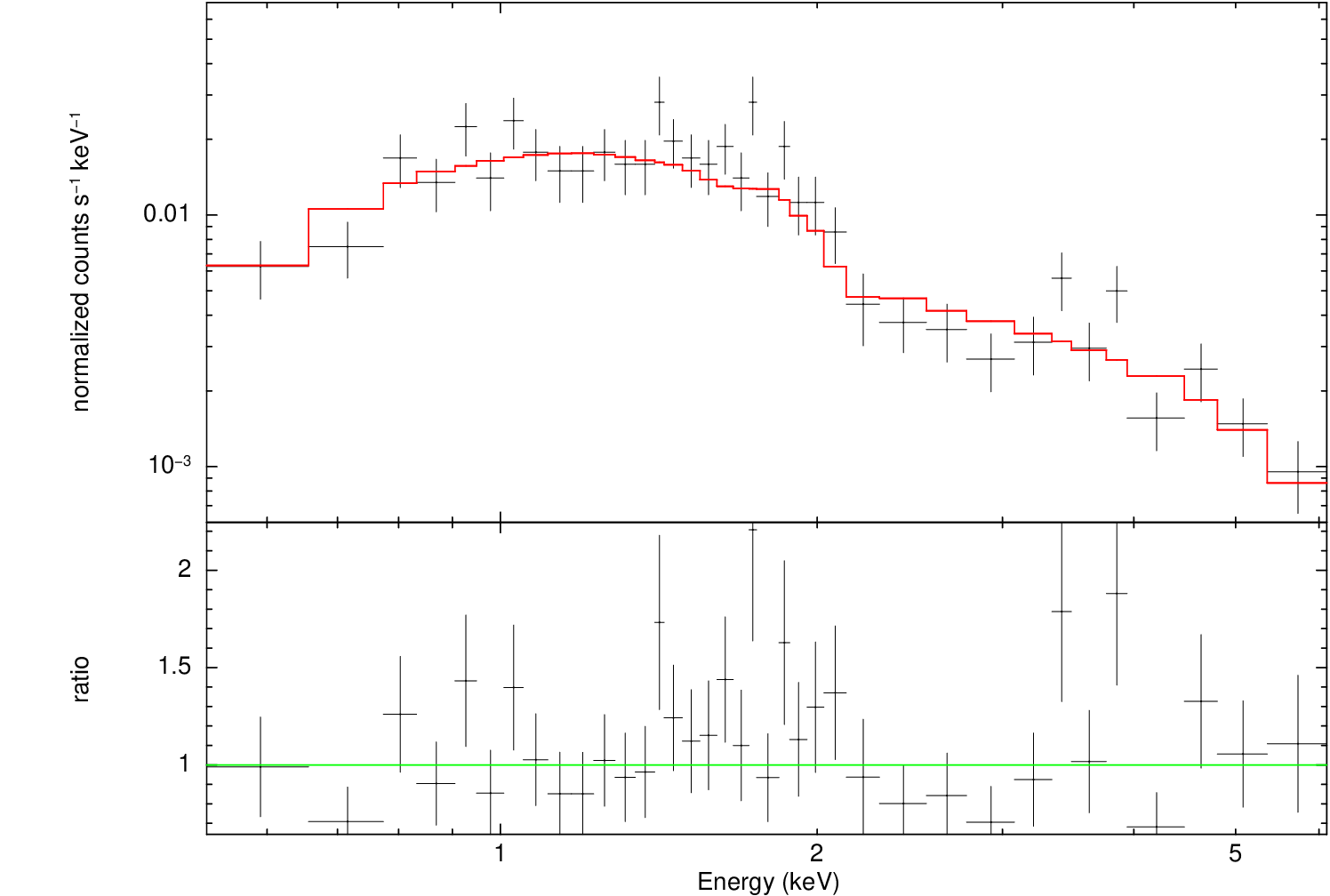}}\quad
\subfigure{\includegraphics[width=0.32\hsize, angle=270]{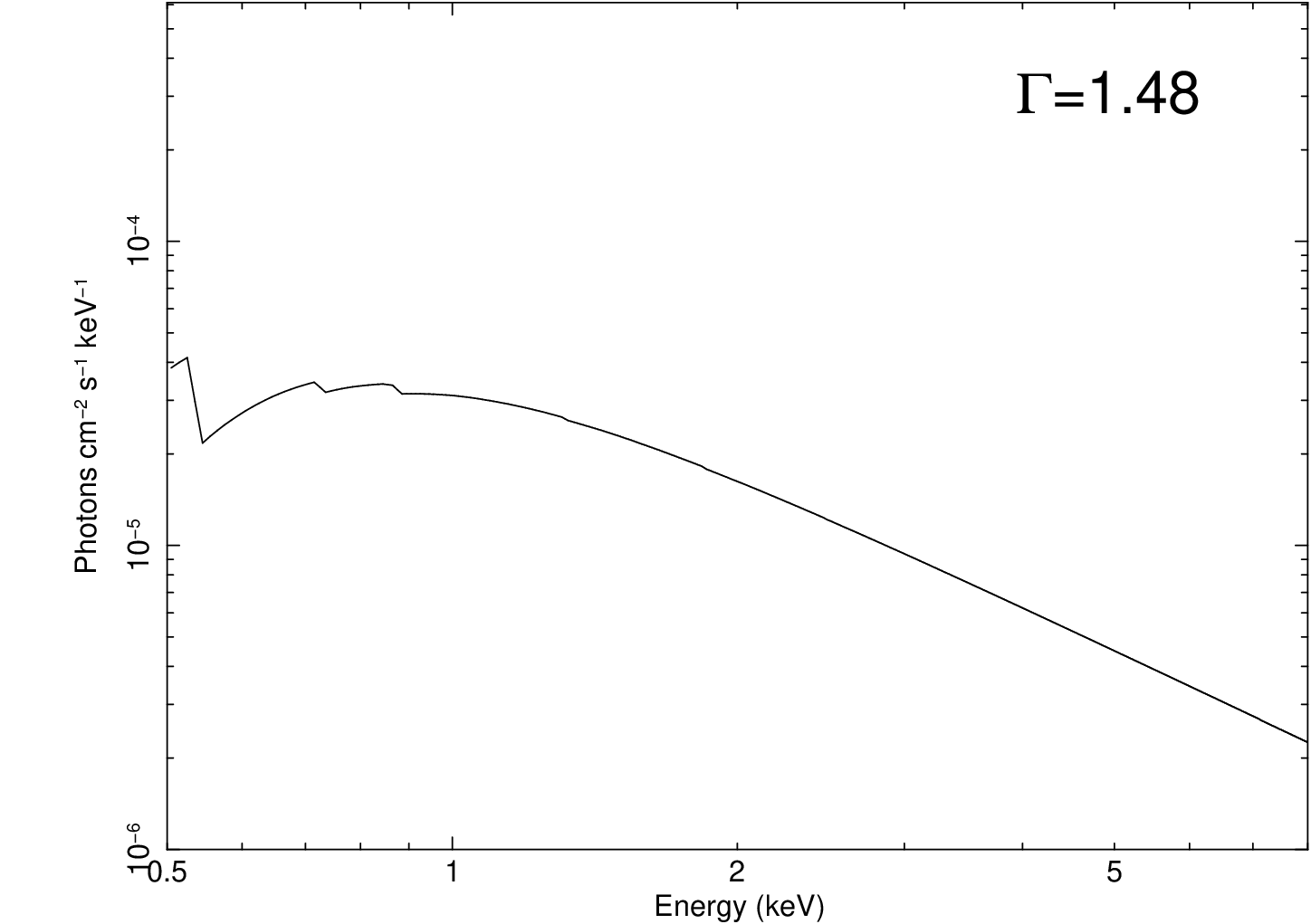}}\quad
\subfigure{\includegraphics[width=0.32\hsize, angle=270]{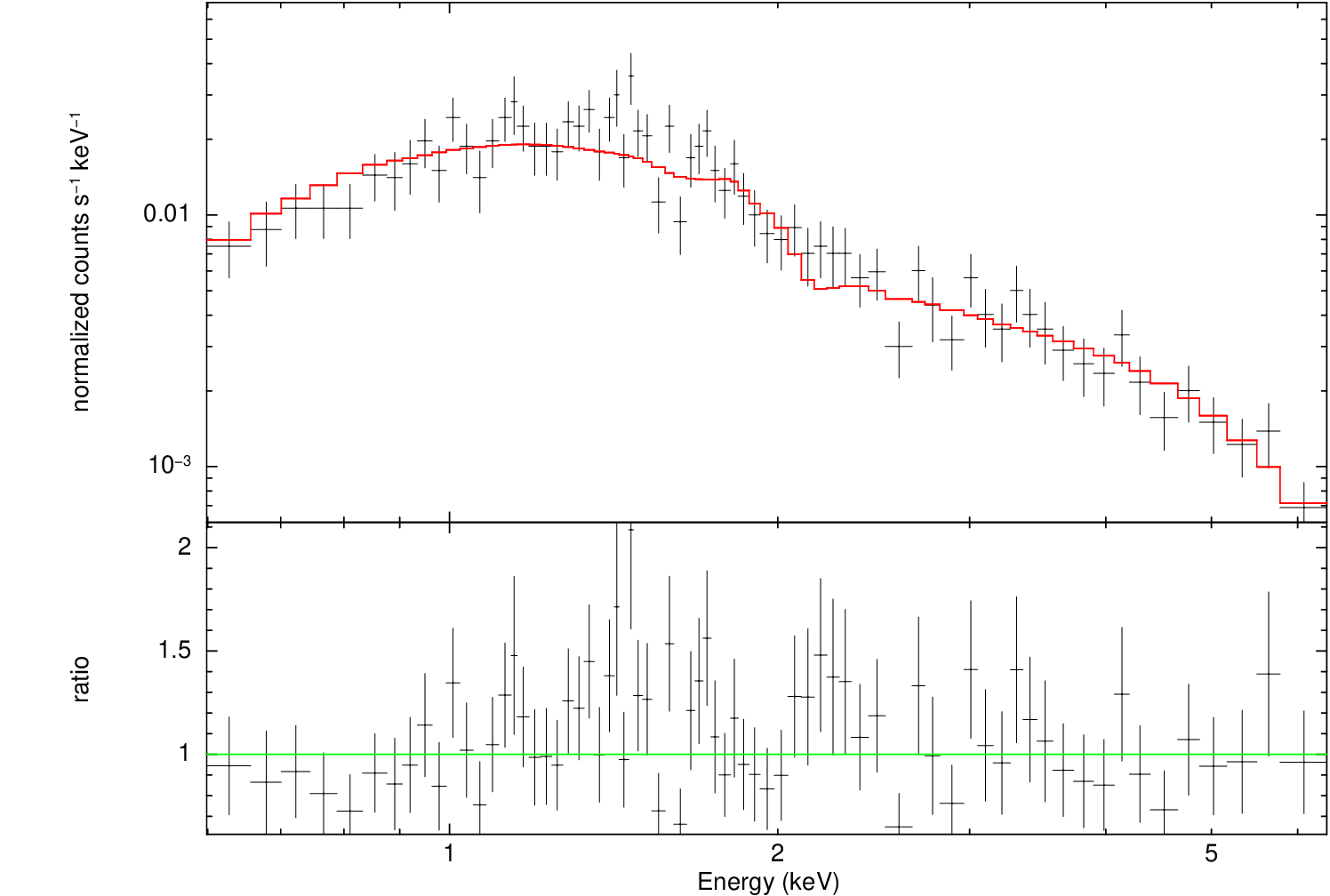}}\quad
\subfigure{\includegraphics[width=0.32\hsize, angle=270]{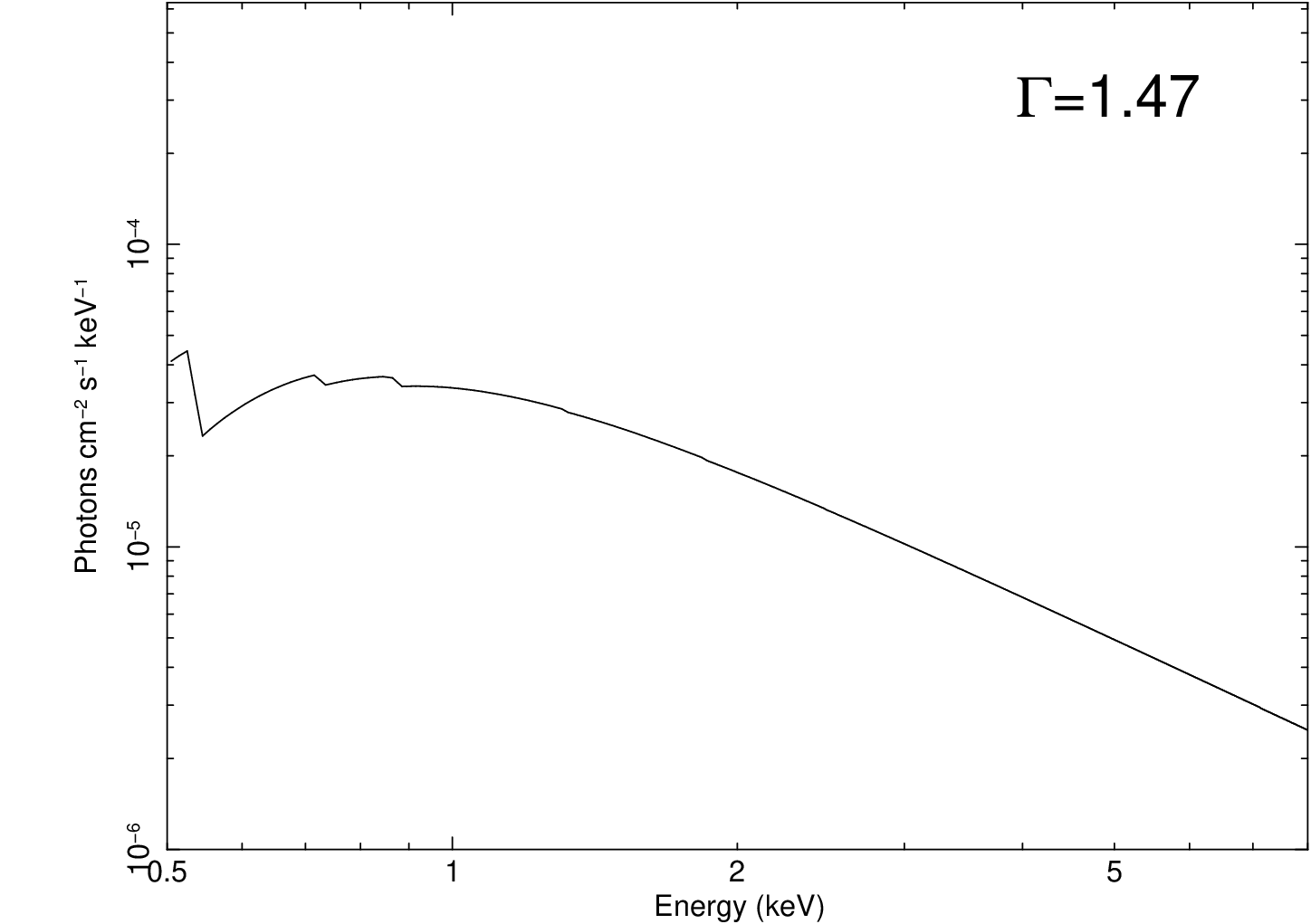}}\quad
\caption[b]{\small (Top) Data and model of observation 4451.  (Middle) Data and
model of observation 4452.  (Bottom) Data and model of observation 4453.  The 
data are on the left, and the models are on the right.  Plotted below the data 
are the data to model ratios.}
  \vspace{0cm}
  \label{fig:samplespectra}
\end{figure*}

\begin{table*}[htb]
\caption[t]{Chandra Observations of V4641 Sgr: Best-Fit Spectral Parameters}
\label{tab:par}
\begin{center}
\begin{tabular}{lccccccccr}
\tableline
Observation & MJD & $\phi$ & $\Gamma$ & $f_{refl}$ & K & $N_{H}$ & $f_{\rm cov}$ & $\log{L/L_{\rm Edd}}$ & $\chi^{2}~({\rm DOF})$\\
~ & ~ & ~ & ~ & ~ & $\times10^{-5}$ & $10^{22}~{\rm cm}^{-2}$ & ~ & ~ & ~ \\
\tableline
4451 & 53203.45 & $0.97\pm0.01$ & $1.96\pm0.08$ & $3.7^{+1.4}_{-1.2}$ & $390\pm20$ & $48\pm4$ & $0.97^{+0.01}_{-0.02}$ & -3.7 & 37.27 (47) \\[1ex]
4452 & 53216.76 & $0.69\pm0.01$ & $1.48^{+0.19}_{-0.12}$ & $0.0002^{+6}_{-3}$ & $4.9\pm0.5$ & --- & --- & -5.5 & 38.32 (34) \\[1ex]
4453 & 53227.18 & $0.39\pm0.01$ & $1.47^{+0.09}_{-0.08}$ & $0.0001^{+2}_{-1}$ & $5.3\pm0.3$ & --- & --- & -5.4 & 65.76 (68)\\[1ex]
\tableline
\end{tabular}
\end{center}
\tablecomments{Parameters of V4641 found for the observations in our sample.  
Columns 3-9 are the best fit spectral parameters found using the model 
$tbabs~\times~pcfabs~\times~pexmon$.  For observations 4452 and 4453, 
{\it pcfabs} was not included in the model.  The uncertainties are the 90\% 
statistical uncertainty calculated using the ``error'' command in XSPEC.  The 
Eddington fractions were estimated from the total unabsorbed flux between 
0.5-8.0 keV, and assuming $M=9.6~M_{\odot}$ and $d=9.6~{\rm kpc}$.  The phases 
($\phi$) were calculated using $T_{0}={\rm HJD}~2447707.4865\pm0.0038 $ and 
$P_{\rm orb}=2.81730\pm0.00001~{\rm days}$ (Orosz et al.\ 2001).}
\vspace{0.5cm}
\end{table*}

\vspace{1.0cm}

The data and corresponding models are plotted in
Figure~\ref{fig:samplespectra}, and the results are listed in Table 1.
For observation 4451, the internal column is measured to be $N_{H} =
4.8(4) \times 10^{23}~{\rm cm}^{-2}$, commensurate with some columns
within Seyfert-2 AGN (e.g. Matt et al.\ 2000; Risaliti, Elvis, \&
Nicastro 2002), and short of being Compton--thick.  Because much of
the incident spectrum is obscured, the reflection fraction is found to
be much higher than unity; indeed, as per Seyfert-2 spectra, it is
reflection--dominated.  An important outcome of this spectral model is
that a fairly typical incident power-law index is recovered: $\Gamma =
1.96(8)$.  Both this model, and the model proposed by Gallo, Plotkin,
\& Jonker (2014), achieve excellent fits ($\chi^{2}/\nu \leq 1$); the models
cannot be distinguished on statistical grounds using these data.

The second and third spectra in the sequence, observations 4452 and
4453, do not require any absorption beyond the Galactic column, nor do
they require significant reflection.  They are well described with a
simple power-law model, typical of black holes at low Eddington
fraction.  Apart from the lack of an absorbing column, these two 
observations differ from the first in several other respects.  The
power-law index is softer in the first observation ($\Gamma\approx
1.96$, see Table 1).  Moreover, the unabsorbed flux in observation
4451 is a factor $\gtrsim50$ higher than in observations 4452 and 4453.
In tandem, then, these facts suggest a degree of continuum evolution
as the outburst declined to a quiescent flux level.

\section{Discussion}
We have reanalyzed three archival {\it Chandra} spectra of V4641~Sgr,
obtained as the source declined into a quiescent state.  The first
spectrum in the sequence (observation 4451; see Table 1 and Figures 1
and 2) is of particular interest, as it shows strong deviations from a
simple power-law.  This spectrum is very
similar to those commonly obtained from Seyfert-2 AGN (e.g. Turner et
al.\ 1997; Matt et al.\ 2000; Risaliti, Elvis, \& Nicastro 2002;
Bianchi et al.\ 2009).  When fit with a model based on Seyfert-2
spectra -- one including partial covering absorption and distant
reflection -- an excellent fit is obtained, and a typical incident
spectrum is recovered.  In this section, we examine ways by which this
spectrum might arise in V4641~Sgr, implications for accretion flows
onto black holes, and means of testing this model.

As briefly mentioned above, V4641~Sgr is not a standard X-ray binary.
It is the only transient high mass X-ray binary with a
dynamically--confirmed black hole primary (Orosz et al.\ 2001).  Our
expectations of high mass X-ray binaries harboring black holes are
built upon persistent (but variable) sources such as Cygnus X-1, LMC
X-1, and LMC X-3.  The phenomena possible when black holes enter and
return to quiescence in the presence of a high-mass companion are
unexplored apart from V4641~Sgr.  Although the companion wind in
V4641~Sgr is likely to be weaker than in a source like Vela X-1,
accretion-induced changes in photoionization and shocks in a companion
wind may cause inhomogeneities within it, and could lead to a complex
absorption geometry (see, e.g., Watanabe et al.\ 2006).  Of course, if
the wind is clumpy, then a complex absorption geometry would also be
likely.

Recently MacDonald et al. (2014) revisited the orbital parameters of
V4641 Sgr by examining photometric data obtained over the last decade.
They found that the companion star to V4641 Sgr exists in two states:
a ``passive'' state in which the secondary exhibits very little
intrinsic variability and the variations in the optical light curve
are caused by the orbital modulation, and an ``active'' state, in
which variability in the light curve is dominated by the intrinsic
variability in the secondary.  Based on MacDonald et al.\ (2014) and
prior work by Orosz et al.\ (2001), we find that observation 4451
occurs during the active state of the secondary, and observations 4452
and 4453 occur in the passive state.  That observations 4452 and 4453
are so different from observation 4451 suggests that the peculiar
spectrum may be partially caused by the increased/changed wind geometry etc. 
from the secondary.  An active X-ray state combined with an active companion
(wind) state could cause the shocks and fronts that would give rise to
the observed spectrum, aided or unaided by clumps.

In a study of Seyfert galaxies, Akylas \& Georgantopoulos (2009) found
that the lower luminosity sources more often had unabsorbed nuclei,
while the higher luminosity sources often had internal hydrogen column
densities as high as $\approx10^{24}~{\rm cm}^{-2}$.  The unabsorbed
sources had Eddington-scaled luminosities comparable to observations
4452 and 4453, while observation 4451 had an Eddington-scaled
luminosity closer to that of the obscured sources.  That fact that the
absorption behavior observed in V4641~Sgr seems to mirror Seyfert
behavior may simply be coincidence.  However, it could represent a
physical similarity.  One possibility may be a clumpy wind from the
accretion flow operates at the Eddington fraction of observation 4451,
but disappears at lower luminosity.  Elitzur \& Shlosman (2006)
developed such a model for AGN.

V4641~Sgr is also unique in the degree by which the inner accetion
disk appears to be misaligned with the binary system ($i \leq 10^{\circ}$
versus $i = 60-70^{\circ}$; Orosz et al.\ 2001).  Depending on the
vertical profile of the disk, the nature of the inner accretion flow,
and how these geometries vary with the mass accretion rate, it is
possible that the outer disk may act like the torus in AGN, and serve
to partly or nearly entirely obscure the inner disk.  This is another
means by which a Seyfert-2--like geometry and spectrum might arise in
V4641~Sgr; it is independent of the companion wind, but of course the
two could act in concert.


In order to model the spectrum of observation 4451, prior efforts
developed some jet-focused models, including extremely strong
Doppler-split lines from SS-433--like jets and/or peculiar
Comptonization (Gallo, Plotkin, \& Jonker 2014).  If the 4--8 keV
spectrum is due to a set of lines, the lines are far stronger than
those observed in SS~433 (e.g. Marshall et al.\ 2013), even though a
far higher Eddington fraction would be estimated from SS~433 if it were
viewed down its axis.  Indeed, the jets in V4641~Sgr would have a bulk
Lorentz factor of 15--45 (Gallo, Plotkin, \& Jonker 2014).  While
Lorentz factors this large are inferred in jets from supermassive
black holes, they are considerably higher than the Lorentz factors
typically observed in jets originating from other stellar mass black
hole systems in the hard or quiescent states (see for example Fender,
Belloni, \& Gallo 2004).  Moreover, the continuum implied in such fits
is unusually hard, and inconsistent with standard Comptonization (
$\Gamma =$0.93--1.07), whereas our model recovers a more typical
continuum.

It is possible that precessing jets have been observed in V4641~Sgr.
However, in view of the qualitative and quantitative similarities
between observation 4451 and well-known Seyfert-2 spectra, in view of
the likely complexity of the environment in this transient,
mis-aligned, high-mass X-ray binary, and in view of the fact that a
canonical continuum can be recovered with an absorption plus
reflection model, we suggest that the latter is also a compelling
interpretation of the data.


One way to constrain the nature of the absorption is to examine the 
orbital geometry of the system at the times of our observations.
In Cygnus X-1 at least, obscuration does appear to be phase--dependent.
Any phase dependence may help to distinguish how the spectrum obtained
in observation 4451 arose, so we calculated the phase ($\phi$) of all three
observations using orbital period and photometric $T_{0}$ calculated
by Orosz et al. (2001; photometric $T_{0}$ is defined by Orosz et al. 
as the time of the superior conjunction of the secondary).  The phases 
are listed in Table~\ref{tab:par}
along with the uncertainties estimated from uncertainties in their
phase zero-point and orbital period.  MacDonald et al.\ (2014) estimated 
a more recent $T_{0}$ using the same orbital period from Orosz et al.\ 
(2001).  Folding our observations on that $T_{0}$, the phase is 
comparable to that found by Orosz et al. Thus our choices of $T_{0}$ 
and $P_{\rm orb}$ (and, in turn, the phases we compute using them) 
are consistent with more recently performed studies of the secondary.

If we then accept these phase determinations, we find that 
observation 4451 occurs near phase 0.0 (i.e. when the black hole is closest to
us).  This nominally implies that the absorption may not be tied to an
enhancement in the wind that is constant in binary phase, and suggests
that the absorption is more closely tied to increased activity in the
secondary and/or the inner accretion flow, and/or related effects on
the outer accretion disk.

In summary, the unusual spectrum revealed in V4641~Sgr is similar to
spectra obtained in Seyfert-2 AGN, and we suggest that a variety of
plausible scenarios -- some tied to the HMXB nature of V4641~Sgr,
others to its odd system parameters -- could produce the geometry
consistent with a Seyfert-2--like spectral model.  Other origins for the
spectrum -- including X-ray line emission from extreme jets and/or
Comptonization -- cannot be ruled out, but we argue that
partial-covering absorption and distant reflection provide
a familiar explanation.  Future observations that sample
a declining X-ray flux trend in V4641~Sgr, and/or the orbital phase,
may be able to better reveal the nature of the spectrum.

The authors acknowledge helpful and supportive conversations with
Elena Gallo, Richard Plotkin, and Peter Jonker.  We also thank the
anonymous referee for a helpful review of this work.

\end{document}